\documentclass[useAMS,usenatbib,usedcolumn]{mn2e}

\usepackage{graphicx} 
\usepackage{times}

\title[Gigahertz-peaked spectra pulsars in Pulsar Wind Nebulae]
{Gigahertz-peaked spectra pulsars in Pulsar Wind Nebulae}
\author[Basu et al.]{R. Basu$^{1,2}$\thanks{e-mail: rahulbasu.astro@gmail.com}, K. Ro\.zko$^{2}$, J. Kijak$^{2}$,  W. Lewandowski$^{2}$\\
$^{1}$ Inter-University Centre for Astronomy and Astrophysics, Pune, 411007, India \\
$^{2}$ Janusz Gil Institute of Astronomy, University of Zielona G\'ora, ul. Szafrana 2, 65-516 Zielona G\'ora, Poland \\
}
\begin{document}



\maketitle

\label{firstpage}

\begin{abstract}
We have carried out a detailed study of the spectral nature of six pulsars 
surrounded by Pulsar wind nebulae (PWN). The pulsar flux density were estimated
using the interferometric imaging technique of the Giant Metrewave Radio 
Telescope at three frequencies 325 MHz, 610 MHz and 1280 MHz. The spectra 
showed a turnover around gigahertz frequency in four out of six pulsars. It has
been suggested that the gigahertz peaked spectra (GPS) in pulsars arises due to
thermal absorption of the pulsar emission in surrounding medium like PWN, HII 
regions, Supernova remnants, etc. The relatively high incidence of GPS 
behaviour in pulsars surrounded by PWN impart further credence to this view. 
The pulsar J1747$-$2958 associated with the well known Mouse nebula was also 
observed in our sample and exhibited GPS behaviour. The pulsar was detected as 
a point source in the high resolution images. However, the pulsed emission was 
not seen in the phased array mode. It is possible that the pulsed emission was 
affected by extreme scattering causing considerable smearing of the emission at
low radio frequencies. The GPS spectra were modeled using the thermal 
free-free absorption and the estimated absorber properties were largely 
consistent with PWN. The spatial resolution of the images made it unlikely that
the point source associated with J1747$-$2958 was the compact head of the PWN, 
but the synchrotron self-absorption seen in such sources was a better fit to 
the estimated spectral shape.
\end{abstract}


\section{Introduction}
\noindent
The radio emission from pulsars are characterized by a steep power law spectrum 
with typical spectral index around $-$1.6 \citep{lorimer1995,maron2000,
jankowski2017} over a wide frequency range from 0.1 -- 10 GHz. A new spectral 
type in radio pulsars was identified by \citet{kijak2007} where the pulsar 
spectra were observed to show a turn over around gigahertz frequencies. These 
sources are called gigahertz-peaked spectrum (GPS) pulsars and only a handful 
of such sources are known. A systematic search for GPS pulsars have been 
conducted since the initial discovery \citep{kijak2011a,kijak2011b,kijak2013,
dembska2014,dembska2015a,dembska2015b,basu2016,kijak2017,jankowski2017} with 
seventeen pulsars confirmed to exhibit GPS behaviour. The GPS pulsars are 
usually associated with young energetic sources which are found in peculiar 
environments like Pulsar wind nebulae (PWN), HII regions, Supernova remnants 
(SNR), etc. This motivated \citet{kijak2011b,lewandowski2015a,rajwade2016} to 
suggest the thermal absorption model based on the initial ideas of 
\citet{Sieber1973}, where the pulsar emission is absorbed in the specialized 
environments, as a likely candidate for GPS behaviour in pulsars. The model has
been subsequently used by \citet{basu2016,kijak2017} to successfully explain 
the spectral behaviour in a majority of GPS pulsars. The environmental effect 
on the pulsar spectrum is further highlighted by the observation that out of 
seventeen known GPS pulsars four pulsars are associated with PWN. This high 
incidence of GPS behaviour near PWNs make them likely candidates for potential 
exploration of GPS behaviour. 

The GPS pulsars based on their peculiar environments are usually associated 
with relatively high dispersion measures with values usually greater than 
200~pc~cm$^{-3}$. This further implies that these sources are affected by 
scattering \citep{lewandowski2013,lewandowski2015b}. In such cases the standard
pulsar observation techniques results in serious underestimation of the pulsar 
flux density due to the presence of long scattering tails. \citet{dembska2015a}
showed that interferometric imaging provides the only possible way to securely 
measure the flux density in highly scattered pulsar profiles. Additionally, 
\citet{basu2016} also demonstrated that the interferometric techniques were 
vastly superior to the standard pulsar flux density measurements from a 
phased-array owing to the possibility of self calibration. In this work we 
attempt to look for GPS behaviour in six pulsars with associated PWN by 
characterising their spectral nature around gigahertz frequencies. We have 
employed the interferometric studies to measure the pulsar flux density at each
observing frequency. In addition a detailed modeling using the thermal 
absorption model has been carried in each case where a GPS behaviour is 
identified.

\section{Observing details}
\noindent
The observations were conducted using the Giant Metrewave Radio Telescope 
(GMRT) located near Pune, India. The observations were primarily carried out 
using standard interferometric schemes with strategically spaced calibrators 
interspersed with the sources as detailed in \citet{dembska2015a}. In one 
pulsar J1747$-$2958 we observed simultaneously in the phased-array mode along 
with the interferometric observations \citep{basu2016} to look for pulsed 
emission. The sources were observed at three different frequencies 325 MHz, 610
MHz and 1280 MHz to characterize the low radio frequency spectra and search 
for GPS behaviour. The 610 MHz observations were conducted in August 2015, 
while the 325 MHz and 1280 MHz observations were carried out between November
and December 2016. Additionally, the pulsar J1747$-$2958 was also observed
separately between April to May 2017 at all three frequencies to check the 
variability in the measured flux density. We observed each pulsar for 
approximately one hour duration. The flux calibrators 3C286 and 3C48 were 
observed during each observing run to calibrate the flux density scale. In 
addition the two phase calibrators 1714$-$252 and 1822$-$096 were observed at 
regular intervals to correct for the temporal variations as well as 
fluctuations in the frequency band. The flux scales of 3C48 and 3C286 were set 
using the estimates of \citet{perley2013} and used to calculate the flux 
density of the different phase calibrators during each observing session as 
shown in Table \ref{tabobs}. The observing mode during each session is 
also shown in the table with IA corresponding to interferometric array and PA 
to phased-array observations. The imaging analysis was carried out using the 
Astronomical Image Processing System (AIPS) similar to \citet{dembska2015a,
basu2016}. Initially all the six sources were observed at 610 MHz to look for 
GPS nature. In The subsequent observations at 325 MHz and 1280 MHz we only 
followed up on the four sources which indicated turnover in the spectra. 

\begin{table}
\resizebox{\hsize}{!}{
\begin{minipage}{80mm}
\caption{Observing details.}
\centering
\begin{tabular}{cc D{,}{\pm}{3.3}c}
\hline
 & & & \\
Obs Date & Phase Calibrator & \multicolumn{1}{c}{Calibrator Flux} & Obs Mode\\
 &  & \multicolumn{1}{c}{({\footnotesize Jy})} & \\
\hline
\multicolumn{3}{c}{\underline{325 MHz}} &  \\
 12 November, 2016 & 1714$-$252 & 5.1,0.4 & IA \\
 & & & \\
 26 November, 2016 & 1714$-$252 & 5.0,0.3 & IA \\
 & & & \\
 30 April, 2017 & 1714$-$252 & 5.0,0.3 & IA,PA \\
 & & & \\
 27 May, 2017 & 1714$-$252 & 5.0,0.3 & IA,PA \\
 & & & \\
\multicolumn{3}{c}{\underline{610 MHz}} & \\
 15 August, 2015 & 1714$-$252 & 4.7,0.3 & IA \\
                 & 1822$-$096 & 6.1,0.4 & \\
 & & & \\
 29 August, 2015 & 1714$-$252 & 4.5,0.3 & IA \\
                 & 1822$-$096 & 6.0,0.4 & \\
 & & & \\
 10 May, 2017 & 1714$-$252 & 4.8,0.3 & IA,PA \\
 & & & \\
 27 May, 2017 & 1714$-$252 & 5.2,0.3 & IA,PA \\
 & & & \\
\multicolumn{3}{c}{\underline{1280 MHz}} & \\
 14 November, 2016 & 1714$-$252 & 2.5,0.2 & IA \\
 & & & \\
 11 December, 2016 & 1714$-$252 & 1.7,0.1 & IA \\
 & & & \\
 28 April, 2017 & 1714$-$252 & 2.4,0.2 & IA,PA \\
\hline
\end{tabular}
\label{tabobs}
\end{minipage}
}
\end{table}

\section{Results}
\begin{figure*}
{\mbox{\includegraphics[scale=0.63,angle=-90.]{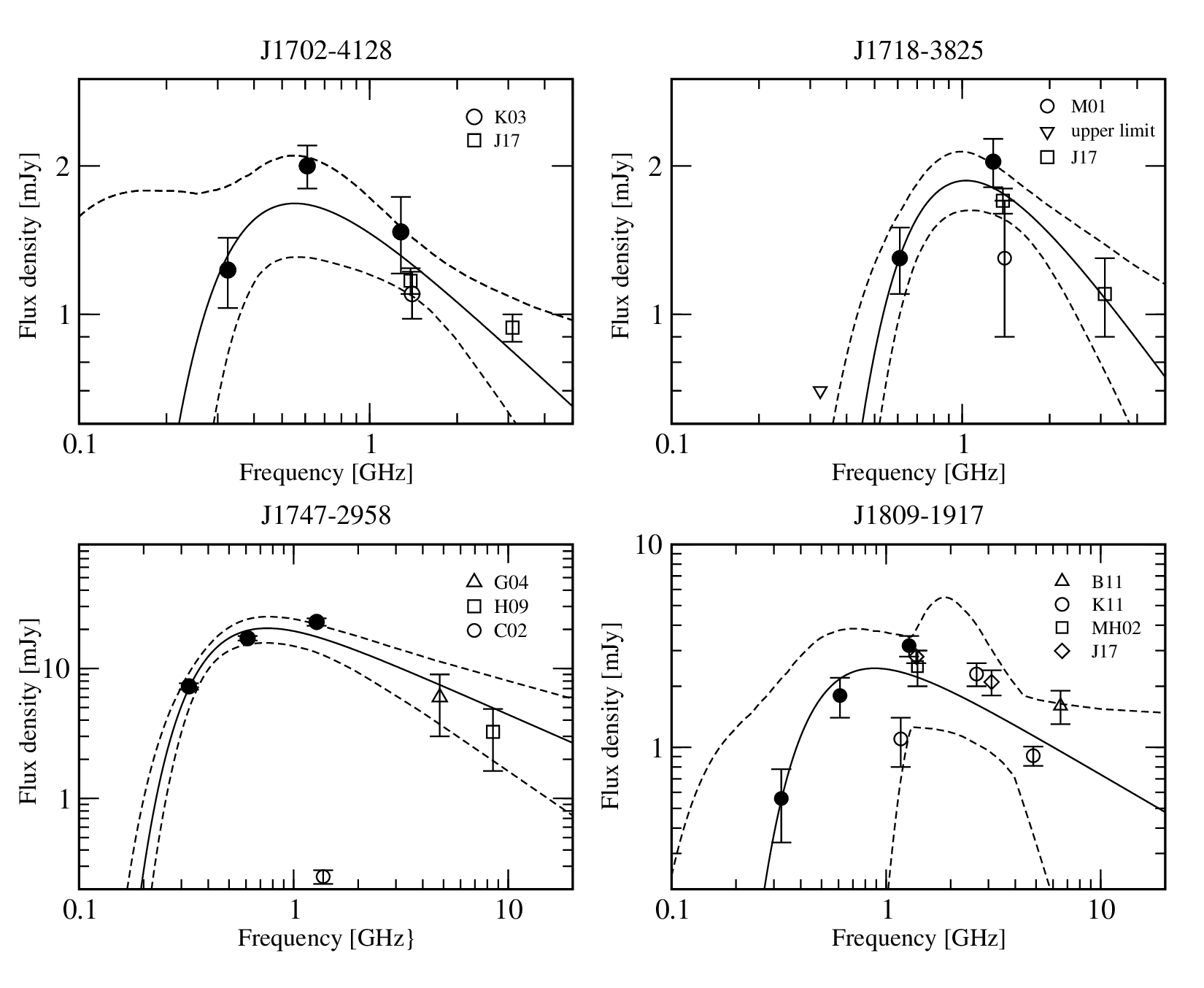}}} 
\caption{The spectra of the four pulsars J1702$-$4128 (top left), J1718$-$3825
(top right), J1747$-$2958 (bottom left) and J1809$-$1917 (bottom right). The 
filled circles represent our measurements reported in this paper while the
open symbols represent flux density measurements from earlier works, 
K03 - \citet{kramer2003}, M01 - \citet{manchester2001}, 
G04 - \citet{gaensler2004}, H09 - \citet{hales2009}, C02 - \citet{camilo2002},
B11 - \citet{bates2011}, K11 - \citet{kijak2011b}, MH02 - \citet{morris2002}, 
J17 - \citet{jankowski2017}.
In addition the non-detection at 325 MHz for J1718$-$3825 is shown as an upper
limit. The figures also shows the spectral fits using the thermal absorption 
model (see text) along with the 1-$\sigma$ envelopes to the fits. The 1374 MHz 
measurement of \citet{camilo2002} is supposed to be severely underestimated due
to scattering and not used for the spectral fits.}
\label{figspec}
\end{figure*}

\noindent
In Table \ref{tabflux} we report the measured flux density of the pulsars at 
the three frequencies. For two cases J1857+0143 and J1913+1011 the flux density
was measured initially at 610 MHz and indicated a typical power law spectra. As
a result they were not observed subsequently at the other frequencies. In all 
pulsars except J1747$-$2958 the pulsar were seen as a point source without any
indication of the diffuse PWN. The associated PWN in each case were identified 
in X-ray observations \citep{Kargaltsev2010}. The pulsar J1747$-$2958 is 
associated with the bow shock nebulae G359.23-0.82 also known as the Mouse 
nebula \citep{yusef1987,camilo2002}. The extended nebula was seen at each of 
the frequencies. The pulsar was detected as a point source in high resolution 
images as discussed in the next section. In the four pulsars J1702$-$4128, 
J1718$-$3825, J1747$-$2958 and J1809$-$1917 the low frequency flux density 
showed a turnover in the spectra indicating GPS behaviour. A detailed spectral 
fit in each of these four cases was estimated using the thermal absorption 
model as described below. The spectral nature of each pulsar is shown in 
figure~\ref{figspec}.

\begin{figure}
{\mbox{\includegraphics[scale=1.2,angle=-90.]{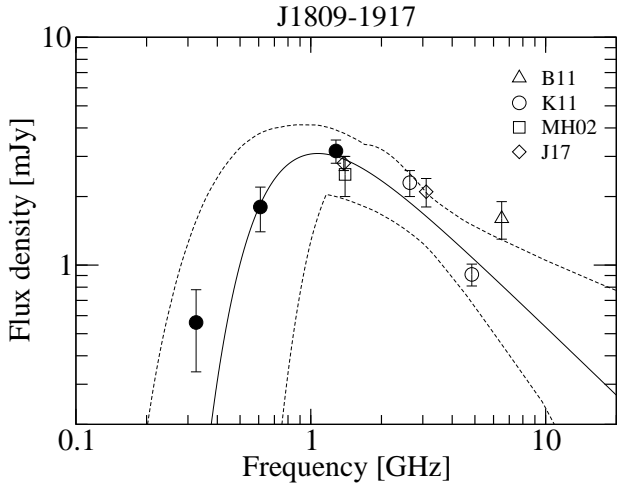}}}
\vskip-3mm
\caption{The figure shows the improved fits of the spectral nature of PSR 
J1809$-$1917 using the thermal absorption model. The fits excluded the 
phased array flux density measurements of \citet{kijak2011b} at 1.2 GHz since 
the data was of poor quality. The improved interferometric measurements at 1280
MHz from our latest observations covered this frequency range for the spectral 
fits.}
\label{figJ1809}
\end{figure}

\begin{table}
\resizebox{\hsize}{!}{
\begin{minipage}{80mm}
\caption{Flux density measurements at the three frequencies 325 MHz, 610 MHz 
and 1280 MHz. The flux density for all sources have been measured twice 
separated by roughly a week and the average value is reported. In case of 
J1747$-$2958 additional measurements were carried out separated by more than a 
month and they are reported separately.}
\centering
\begin{tabular}{cccc}
\hline
 & \multicolumn{3}{c}{Flux (mJy)}\\
 Pulsar & 325 MHz & 610 MHz & 1280 MHz \\
\hline
 J1702$-$4128 & 1.2$\pm$0.2 & ~2.0$\pm$0.2 & ~1.5$\pm$0.3 \\
              &             &              &              \\
 J1718$-$3825 &   $<$~0.7   & ~1.3$\pm$0.2 & ~2.0$\pm$0.2 \\
              &             &              &              \\
 J1747$-$2958 & 7.4$\pm$0.6 & 17.8$\pm$1.0 & 25.2$\pm$1.7 \\
              & 6.7$\pm$0.5 & 15.3$\pm$0.9 & 16.5$\pm$1.2 \\
              & 7.3$\pm$0.6 & 15.2$\pm$0.9 &              \\
              &             &              &              \\
 J1809$-$1917 & 0.6$\pm$0.2 & ~1.8$\pm$0.4 & ~3.2$\pm$0.4 \\
              &             &              &              \\
  J1857+0143  &     ---     & ~2.3$\pm$0.3 &      ---     \\
              &             &              &              \\
  J1913+1011  &     ---     & ~3.4$\pm$0.2 &      ---     \\
\hline
\end{tabular}
\label{tabflux}
\end{minipage}
}
\end{table}

\subsection{Thermal absorption of Pulsar emission}
\begin{table*}
\caption{Estimating the fitting parameters for the Gigahertz-peaked spectra 
using the thermal absorption model}
\centering
\begin{tabular}{cccccc}
\hline
 Pulsar & $A$ & $B$ & $\alpha$ & $\chi^2$ &  $\nu_{\mathrm{p}}$ \\
        &     & K$^{-1.35}$~pc~cm$^{-6}$ &          &          &         (GHz)       \\ 
\hline
        &     &     &          &          &          \\ 
J1702$-$4128 & $0.45^{+0.40}_{-0.24}$ & $0.075^{+0.080}_{-0.072}$ & $-0.55^{+0.36}_{-0.39}$ & $2.30$ & $0.55$ \\
        &     &     &          &          &          \\ 
J1718$-$3825 & $0.43^{+0.46}_{-0.26}$ & $0.42^{+0.24}_{-0.21}$ & $-0.82^{+0.44}_{-0.54}$ & $1.22$ & $1.03$ \\
        &     &     &          &          &          \\ 
J1747$-$2958 & $3.3^{+1.6}_{-1.5}$ & $0.203^{+0.066}_{-0.058}$ & $-0.85^{+0.23}_{-0.28}$ & $4.45$ & $0.73$ \\
        &     &     &          &          &          \\ 
J1809$-$1917 & $0.54^{+0.51}_{-0.30}$ & $0.55^{+2.42}_{-0.43}$ & $-1.00^{+0.57}_{-0.98}$ & $3.53$ & $1.07$ \\
        &     &     &          &          &          \\
\hline
\end{tabular}
\label{tabfit}
\end{table*}

\noindent
The thermal free-free absorption has been used to model the GPS behaviour by 
\citet{lewandowski2015a}. This method was further demonstrated in the 
subsequent works of \citet{basu2016,kijak2017} to explain the spectral 
behaviour in a majority of GPS pulsars. The unabsorbed intensity is expressed 
as a simple power law $I_{\nu}$ = $A(\nu/\nu_0)^{\alpha}$ with the observed 
spectrum given as :
\begin{equation}\label{eqfit}
S(\nu) = \mathrm{A} ~ \left( \frac{\nu}{10 \mathrm{GHz}}\right)^{\alpha} ~ e^{-\mathrm{B}~\nu^{-2.1}},
\end{equation}
where $A$ is the amplitude (i.e. the pulsar intrinsic flux density at 10~GHz), 
$\alpha$ the intrinsic spectral index of the pulsar and $\mathrm{B} = 0.08235 
\times T_{\mathrm{e}}^{-1.35}$EM, EM being the emission measure. The best fit 
to the measured spectra were obtained using the Levenberg-Marquardt least 
squares algorithm, with $A$, $\alpha$ and $B$ as free parameters of the fit. 
Finally, $\chi^2$ mapping was used to estimate the errors in the fitting 
parameters. Table~\ref{tabfit} shows the details of the fitting process in the 
four pulsars. The table also shows the estimated peak frequency, 
$\nu_{\mathrm{p}}$, of the spectra from the fits. In case of PSR J1718$-$3825 
the source was not detected at 325 MHz (only upper limit established) and there
were only three measurements for fitting the three parameter model. We assumed 
an intrinsic spectral index $\alpha$ = $-1.6$, which is the average 
spectral index in the pulsar population \citep{lorimer1995,maron2000,
jankowski2017}, and estimated the other parameters as shown in 
Table~\ref{tabfit}. In \citet{kijak2017} the spectral nature of the pulsar 
J1809$-$1917 was estimated using only high frequency measurements ($>$ 1 GHz). 
We have now carried out lower frequency measurements for this pulsar and 
considerably improved the spectral nature as well as the fitting parameters. 
The resultant spectral fits along with the 1$\sigma$ envelopes are also shown 
in figure~\ref{figspec}. In the case of PSR J1809$-$1917 it was found that
the flux measurements at 1.2 GHz \citep{kijak2011b} using the phased array mode
was from a noisy profile with weak signal \citep[the profile is shown in 
Appendix of][]{lewandowski2013}. This resulted in lower values of the flux 
density and considerably degraded the fitting estimates. Our current 
interferometric measurements at 1280 MHz provided coverage at this frequency 
range. In figure \ref{figJ1809} we show improved fits to the spectral 
parameters after excluding the 1.2 GHz values which are reported in the tables.

\section{PSR J1747$-$2958 associated with Mouse Nebula}
\begin{figure*}
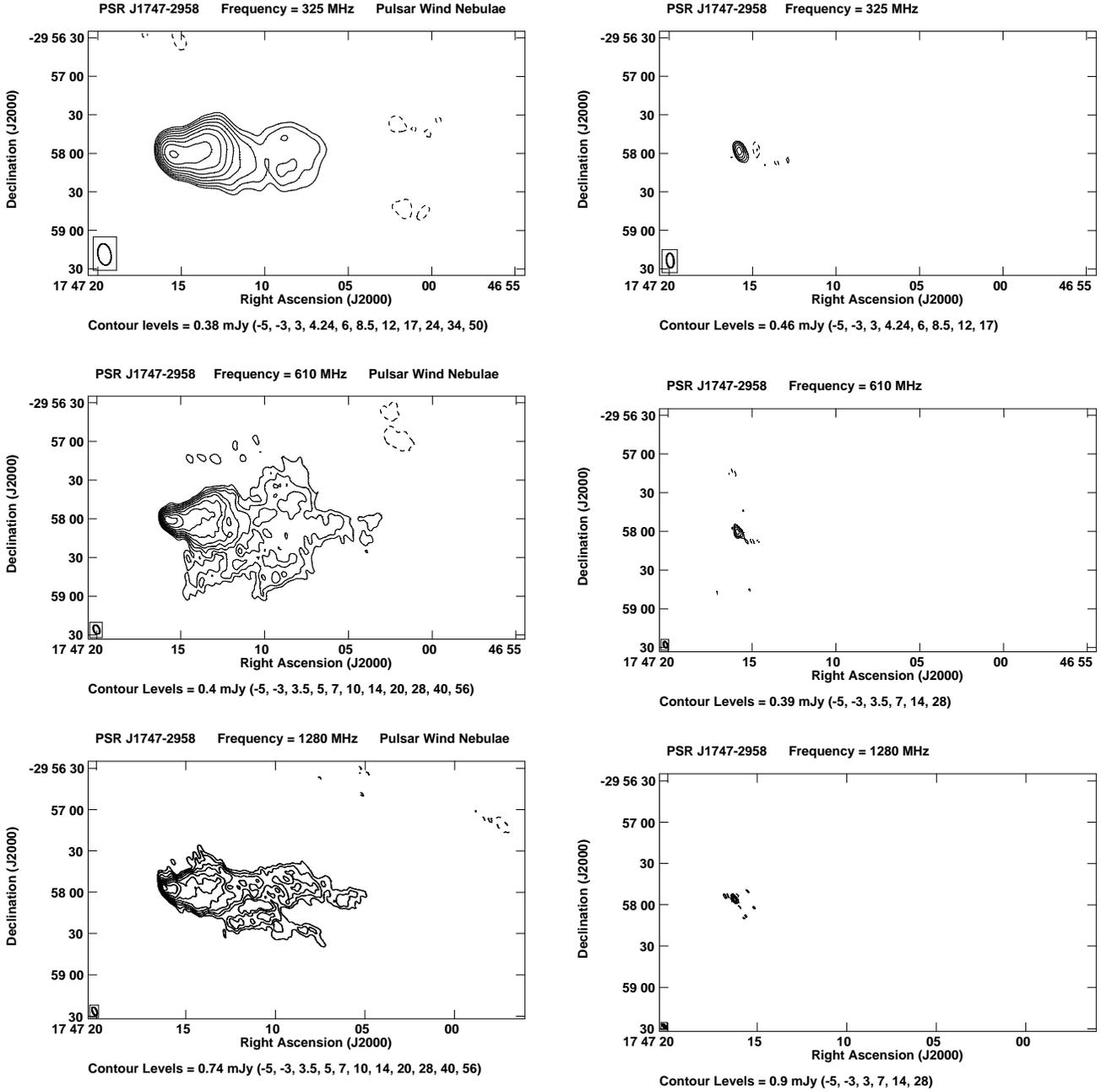

\begin{tabular}{@{}lr@{}}
{\mbox{\includegraphics[scale=0.45,angle=0.]{figure3a.ps}}} &
{\mbox{\includegraphics[scale=0.45,angle=0.]{figure3b.ps}}} \\
{\mbox{\includegraphics[scale=0.45,angle=0.]{figure3c.ps}}} &
{\mbox{\includegraphics[scale=0.45,angle=0.]{figure3d.ps}}} \\
{\mbox{\includegraphics[scale=0.45,angle=0.]{figure3e.ps}}} &
{\mbox{\includegraphics[scale=0.45,angle=0.]{figure3f.ps}}} \\
\end{tabular}
\caption{PSR J1747$-$2958: Pulsar Wind Nebulae (left panel), pulsar point
source (right panel); 325 MHz - top panel; 610 MHz - middle panel; 1280 MHz - 
bottom panel.}
\label{figpwn}
\end{figure*}

\noindent
As discussed in the previous section the standard imaging schemes revealed the
detailed PWN associated with PSR J1747$-$2958 as shown in figure \ref{figpwn} 
(left panel). In order to detect the pulsar we carried out high resolution 
maps for this source. The interferometric imaging involves measuring correlated
signals between antenna pairs recorded in a co-ordinate system known as the 
$uv$-plane. The points in the $uv$-plane, also known as baselines, corresponds 
to projected distances between the antenna pairs and records signals from the 
sky in the fourier domain. The standard imaging technique involves 
reconstructing the relevant part of the sky by carrying out properly weighted 
fourier transforms of the signals recorded in the $uv$ domain. Hence, the scales
of structure in the sky plane are inversely associated with the baselines in 
the $uv$-plane, i.e. the large scale diffuse structures are captured by the 
short baselines while the point sources are more prominent at longest 
baselines. The high resolution images involve using only the longest 
baselines in the $uv$-plane to reconstruct the map of the observed sky. This 
would ensure that the extended sources are resolved out and only the point 
sources in the sky are seen in the images. The results of the high resolution 
maps for the pulsar J1747$-$2958 are also shown in figure \ref{figpwn} (right 
panel). The synthesized beam (representing the telescope resolution) is also 
shown in each case. It is clearly seen that in the high resolution maps the 
diffuse emission is resolved out leaving behind the pulsar which appear as a 
point source having the same size as the synthesized beam. It can be argued 
that some amount of the PWN is left behind in the high resolution maps 
contaminating the pulsar flux density. It should be noted that the measured 
flux density of the pulsar is lowest at 325 MHz and highest at 1280 MHz (see 
Table~\ref{tabflux}). However, the synthesized beam size is highest at 325 MHz,
around 8 arcseconds, and decreases with increasing frequency, around 4 
arcseconds at 610 MHz and 2 arcseconds at 1280 MHz. If the diffuse emission 
contributes to the measured flux density then owing to the largest beam size at
325 MHz the measured flux density should also be highest at this frequency and 
decrease with increasing frequency which is contrary to observations. This 
ensures that the measured flux density from the high resolution maps correspond
to the pulsar J1747$-$2958 which is an unresolved point source. We also tried 
to carry out high resolution maps from the archival data at 4.8 GHz 
\citep{gaensler2004} and 8.5 GHz \citep{hales2009} observed using the Very 
Large Array (VLA). However, these observations were conducted during the more 
compact array configurations of VLA (C and D configurations). This implied that
the $uv$ coverage was sparse at the longest baselines making such high 
resolution studies untenable. For a comparison the maximum baseline length
is 0.6 kilometers in the VLA D configuration which corresponds to lengths of
10-15 k$\lambda$ at these observing frequencies, while the longest baseline 
length for GMRT at the L-band is around 125 k$\lambda$. We have used the peak 
flux density values with a 50\% error bar as an approximation for the pulsar 
flux density to estimate the high frequency spectra for the fitting purpose.

The pulsar J1747$-$2958 has been discovered by \citet{camilo2002} at 1374 MHz 
using the Parkes radio telescope. They reported the measured flux density to be
0.25$\pm$0.03 mJy. However, they did not detect the pulsar in a pulsar-gated 
radio imaging with the Australia Telescope Compact Array (ATCA). The flux 
density measurements are significantly lower than the our estimated values. It 
is most likely that the pulsar emission is affected by significant scattering. 
This would modify the pulsar emission and result in an extended scattering 
tail. As a result the pulsar flux density will be spilled over the profile 
baseline and considerably reduce the estimated flux density using standard 
pulsar observations \citep{dembska2015a}. It should also be noted that gating 
observations which involve ``on-pulse minus off-pulse'' images are also not 
adequate for such studies since the flux density is spilled over in the 
off-pulse window and gets subtracted. In our later observations carried out 
between April-May 2017, we simultaneously observed in the phased array mode and
the interferometric mode of GMRT \citep{basu2016}. But the pulsar profile was 
not detected in the phased array observations. If the pulsar is indeed 
scattered then this is understandable since the peak flux density would be 
below our detection levels (noise rms of 0.1-0.2 mJy at 1280 MHz). It should be
noted that scattering characteristic decay time ($\tau_d$) typically scales as 
$\tau_d \propto \nu^{-4.4}$ with frequency \citep{lewandowski2013}, which makes
lower frequency detections impossible. Higher frequency observations greater 
than 2 GHz would be necessary to characterize the scattering in this pulsar.

The fact, that we could not detect the pulsed emission from PSR~J1747$-$2958, 
and the above mentioned low flux density value of the pulsar measured by
\citet{camilo2002} suggest another possibility. The unresolved point source 
detected in our maps at the tip of the Mouse nebula might also correspond to a 
compact emission from the PWN. The minimum size of the emission is $\sim$2
arcseconds (at the highest frequency). The estimated distance to the nebulae is 
between 2-5 kiloparsecs \citep{camilo2002,gaensler2004}. This corresponds to 
the size of the point source to be around 0.02-0.05 parsec. The typical size of
the comet's head, where the shock is located, in  a comet-shaped bow-shock PWNe
is of the order of 1~parsec \citep[see for example][]{bucciantini2002}, which 
is well beyond the estimated size of the point source. But at the moment we 
cannot conclusively exclude the possibility of a very compact PWN, or a ``hot 
spot'' inside it, that could be much smaller. 

If the emission originates from a compact nebula the synchrotron 
self-absorption mechanism in the relativistic pulsar wind becomes important. 
This will also lead to a likely turnover at the low frequency regime. To 
explore this possibility we modeled the spectra \citep[excluding the 
measurements of][]{camilo2002} using a synchrotron self-absorption spectrum 
with the functional form \citep{izvekova1981}: 
\begin{equation}
\label{synchr_fit}
S_{\nu} = b \left(\frac{\nu}{\nu_0}\right)^{\alpha} \exp \left\{\frac{\alpha}{\beta} \left(\frac{\nu}{\nu_c}\right)^{-\beta}\right\}, \
\end{equation}
\noindent
where $\nu_0=10$~GHz similar to the earlier thermal absorption fits, $\nu_c$ 
being the synchrotron self absorption critical frequency, $b$ is a flux scaling
constant, $\alpha$ is the synchrotron spectral index and $0<\beta\leq2$ is the 
turnover smoothness parameter. 

\begin{figure}
{\mbox{\includegraphics[scale=0.33,angle=-90.]{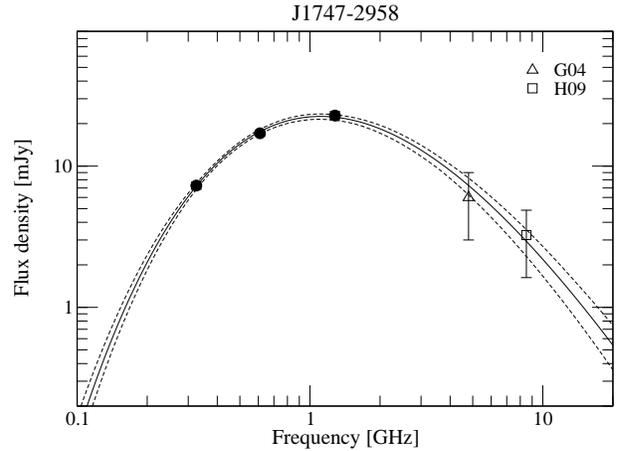}}} 
\vskip-3mm
\caption{The interferometric measurements of the Mouse nebula point source flux
density with the synchrotron self-absorption model fitted. Filled circles 
represent our measurements, G04 - from \citet{gaensler2004},  H09 - from 
\citet{hales2009}.}
\label{spec_synchr}
\end{figure}

The result of the fitting process is shown in Figure~\ref{spec_synchr}. The 
fitting was performed using four free parameters, and the best fit values are: 
normalized $\chi^2=0.18$, $b=58^{+8}_{-7}$~mJy, $\nu_c=1.112\pm0.045$~GHz, 
$\alpha=-3.17\pm0.26$ and $\beta=0.4^{+0.25}_{-0.1}$. The values of $\alpha$ 
and $\beta$ are well within the bounds allowed for the synchrotron 
self-absorption model. As seen in the figure the spectral nature is better 
represented by the fits from this model. However, the size of the point 
source in our maps is much smaller (0.05 parsec) compared to the size of a 
typical PWN (1 parsec). This makes it more likely that despite the better fits
from the synchrotron self absorption model compared to the thermal absorption
model, the source is actually the pulsar. With the possibility of strong 
interstellar scattering discussed above, only the high frequency observations 
($>$ 2 GHz) using single dish or phased-array can help us measure the clean 
pulsar profile and compare with interferometric flux values. This would 
conclusively resolve the origin of the point source seen in the high resolution
maps.

\section{The GPS behaviour in PWN}
\begin{table*}
\caption{The physical properties of absorber leading GPS behaviour}
\centering
\begin{tabular}{ccccccc} 
\hline
 Pulsar & DM & Age & Size & n$_{\mathrm{e}}$ &  EM &  T \\
        & (pc cm$^{-3}$) & $10^4$ (yr) & (pc) & (cm$^{-3}$) & ($10^4$ pc cm$^{-6}$) & (K) \\ \hline
J1702$-$4128 & $367.1\pm0.7$ & 5.51 & ~0.1 & 1835.5$\pm$3.5   &   33.69$\pm$0.13  & $13300^{+10500}_{-9500}$ \\
  &  &  &  &  &  & \\
              &               &      & ~1.0 & 183.55$\pm$0.35  &  3.369$\pm$0.013  & $2420^{+1910}_{-1730}$ \\
  &  &  &  &  &  & \\
              &               &      & 10.0 & 18.355$\pm$0.035 & 0.3369$\pm$0.0013 &  $439^{+347}_{-313}$   \\ 
  &  &  &  &  &  & \\
\hline
 J1718$-$3825 & $247.4 \pm 0.3$ & 8.95 & ~0.1 &  $1237.0\pm1.5$  &   $15.302\pm0.037$  & $2080^{+890}_{-790}$ \\
  &  &  &  &  &  & \\
              &                 &      & ~1.0 & $123.70\pm0.15$  &  $1.5302\pm0.0037$  &  $378^{+162}_{-144}$   \\
  &  &  &  &  &  & \\
              &                 &      & 10.0 & $12.370\pm0.015$ & $0.15302\pm0.00037$ & $69^{+29}_{-26}$ \\
  &  &  &  &  &  & \\
\hline
 J1747$-$2958 & $101.5\pm1.6$ & 2.55 & ~0.1 &  $507.5\pm8.0$  &   $2.576\pm0.081$   &  $960^{+280}_{-260}$ \\
  &  &  &  &  &  & \\
              &               &      & ~1.0 & $50.75\pm0.80$  &  $0.2576\pm0.0081$  &   $175^{+52}_{-46}$  \\
  &  &  &  &  &  & \\
              &               &      & 10.0 & $5.075\pm0.080$ & $0.02576\pm0.00081$ & $31.7^{+9.4}_{-8.4}$ \\ 
  &  &  &  &  &  & \\
\hline
  J1809$-$1917 & $197.1\pm0.4$ & 5.13 & ~0.1 &   $985.5\pm2.0$   &   $9.712\pm 0.039$   & $1210^{+3970}_{-702}$ \\
  &  &  &  &  &  & \\
              &               &      & ~1.0 &   $98.55\pm0.20$  &  $0.9712\pm 0.0039$  &  $221^{+721}_{-127}$  \\
  &  &  &  &  &  & \\
              &               &      & 10.0 & $9.8550\pm 0.020$ & $0.09712\pm 0.00039$ &    $40^{+131}_{-23}$   \\ 
  &  &  &  &  &  & \\ 
\hline
\end{tabular}
\label{tabphy}
\end{table*}

\noindent
There are around 70 known PWN mostly discovered from X-ray observations. 
The X-ray from PWN is emitted due to synchrotron radiation from relativistic 
pulsar winds shocked in the ambient medium \citep{Kargaltsev2010,
Kargaltsev2013}. Amongst these there are around 40 sources where the associated
radio pulsar has been discovered. We have surveyed these radio pulsars and 
found around 20 sources which are in the GMRT declination range. The detailed 
low frequency spectra, around GHz frequencies, using the GMRT have been 
established in ten pulsars. In this work we have studied six pulsars, 
additionally the pulsar J1740+1000 is reported in \citet{kijak2011a,
dembska2014}, B1800$-$21~in \citet{basu2016}, B1823$-$13 and J1856+0245~in 
\citet{kijak2017}. PSR J1856+0245 was not detected at 610 MHz, but the upper 
limits indicate the spectral shape to be much flatter. Only in two pulsars 
reported here, J1857+0143 and J1913+1011, the spectra resemble a typical power 
law. The role of the pulsar environment in their GPS behaviour have been 
highlighted in previous works \citep{kijak2011a,lewandowski2015a}. There are at
present twenty confirmed cases of GPS pulsars out of which seven sources are 
associated with PWN. The preponderance of GPS pulsars associated with PWN 
provides further evidence for this hypothesis. It has been suggested that the 
intrinsic flux density of young and energetic pulsars associated with PWN are 
usually small \citep{camilo2002}. However, as J1747$-$2958 suggests the pulsars
in such systems are usually associated with scattering which would result in 
severe underestimation of the flux density using standard single dish 
measurements \citep{dembska2015b}. In addition the GPS nature would imply that
the low frequency flux density is much lower than expected from a typical power
law spectra. This also makes the detection of pulsars associated with PWN below
GHz frequencies much more difficult. The remaining pulsars with associated PWN 
where the low frequency spectra is not established should provide rich ground 
for discovering more GPS behaviour, with the interferometric studies being the 
preferred mode of flux density measurement.

Finally, we investigate the physical properties of the absorbing medium in
the four GPS pulsars reported here based on spectral fits of the thermal 
absorption model. It should be noted that spectral fits reported in this 
work are affected by the sparse sampling of the low frequency regime. There are
only a few flux density measurements below 1 GHz to suitably constrain the 
parameters of the fit. There is a paucity of the frequency coverage in 
interferometric arrays like GMRT where observations can be carried out over a
short band centered around specific frequencies. This situation will be 
remedied in the future when newer Telescopes and suitable upgrades to existing 
ones will allow a continuous frequency coverage. In such cases the model 
parameters can be constrained to higher accuracy. However, for our present 
purpose we used the estimated parameters as indications of the nature of the 
absorbing medium and explore their properties. As explained in 
\citet{lewandowski2015a,rajwade2016} the free electrons present in the ISM and 
the absorbing medium contribute to both the dispersion and thermal absorption. 
If we assume an absorber of size $s$ and free electron density $n_e$ uniformly 
distributed, the emission measure is given as EM=$n_e^2\times~s$. The 
contribution to the dispersion measure is $\Delta$DM=$n_e\times~s$. As 
discussed in previous works the different quantities are correlated and would 
require additional inputs like size of the absorbing region as well as 
contribution of the electron density of the absorber to the DM. 

In the absence of any measurements of the absorber parameters like their
size, electron density or free electron temperatures we have carried out some 
speculative estimates from the known environment of young pulsars using the 
fitting parameters of the thermal absorption model. We have investigated three 
primary absorbers found around young pulsar, viz., dense filaments of SNR, PWN 
and HII regions. In Table \ref{tabphy} we show the requirements of the physical
properties of the absorber from the fitting parameters of each pulsar. The top 
row corresponds to SNR filaments with typical size of 0.1 parsec, the middle 
row for PWN with sizes of 1 parsec and the bottom row corresponding to HII 
regions of size 10 parsec. We have assumed that contribution of the DM for each
observer to be $\Delta$DM=0.5DM, and using the size $s$= 0.1,1,10 parsec, 
respectively, the electron densities $n_e$ were calculated in Table 
\ref{tabphy}. We further estimated the EM as explained above using $n_e$ and 
$s$. We used the fitting parameter $B$ (see eq.\ref{eqfit}, and the discussion 
below it) to calculate the free electron temperatures ($T_e$) which are also 
reported in the Table. In case of HII regions the estimated $n_e$ are between 5
and 20 $cm^{-3}$ while $T_e$ are between 30 and 450 K. These values make it 
unlikely for the HII region to be potential absorber. It is possible to have 
electron densities between 10 - 100 $cm^{-3}$ in large HII regions but the 
corresponding electron temperatures are much higher between 1000 and 10000 K 
\citep{tsamis2003}. It should be noted that if we reduce the contribution of 
$\Delta$DM to be less than 0.5DM the estimates will be even lower. In the case 
of SNR filaments the required densities are between 500 and 2000 $cm^{-3}$ and 
$T_e$ are between 1000 and 15000 K. These conditions are consistent with 
previously reported measurements in the SNR filaments \citep{koo2007}. It 
should also be noted that for the thermal absorption model to work, the 
location of the absorber need not be in the immediate vicinity of the source 
but merely along the line of sight. But one drawback for the SNR filaments to 
be the likely absorber is that their typical size is very small (0.1 parsec) 
which reduces the probability of their presence along the line of sight to the 
pulsar. Finally, the most probable absorber are the PWN since their presence is
known around these pulsars. The $n_e$ values for the four pulsars in this case 
are between 50 and 200 $cm^{-3}$, and the corresponding $T_e$ are between 150 
and 2500 K. \citet{lewandowski2015a} explored a simplified model of the 
asymmetric bow-shocked PWN, assuming an uniform electron density and 
temperature across the medium, and showed that the GPS behaviour is affected by
the line of sight towards the PWN. For instance the line of sights passing 
through the tail would cover the maximum absorber size and result in GPS 
spectra. However, the spectra would likely resemble a typical power law if the 
line of sight passes towards the head of the bow shock. This is a possible 
explanation for the two pulsars in our sample which show a typical power law 
spectra. The typical power law spectra would also be possible if the PWN shape 
is not asymmetric like in a bow-shock nebula but more spherically symmetric as 
in the younger Crab pulsar. However, the characteristic age of both pulsars 
J1857+0143 and J1913+1011 are similar to the other GPS pulsars studied here. In
case of the mouse nebulae associated with the pulsar J1747$-$2958, the distance
to the source has been estimated to be 2-5 kiloparsecs \citep{camilo2002,
gaensler2004}. The angular size of the nebula in our maps is $\sim$ 1 arcminute
which gives an approximate size of the PWN to be around 1 parsec. In the PWN 
associated with the pulsar B1951+32 the electron density in front of the shock 
front was estimated from emission lines to be 50--100~cm$^{-3}$ 
\citep{hester1989,li2005}. If we assume the above quantities to be typical 
values of the absorber in PWN then the estimated parameters for the four 
pulsars in table~\ref{tabphy} are largely consistent.

\section{Summary}
\noindent
A detailed measurement of the flux density around gigahertz frequencies were 
carried out for six pulsars associated with PWN using interferometric mode of 
GMRT. The low frequency spectra indicated the presence of GPS behaviour in four
out of six pulsars. There are twenty GPS pulsars know at present out of which
seven pulsars are associated with PWN. In the case of the pulsar J1747$-$2958 
associated with the well known Mouse nebula, high resolution imaging resulted
in correct estimation of the pulsar flux density at low frequencies. The 
previous measurement of the pulsar flux density using standard pulsar mode 
observations appears to be seriously underestimated as the pulsar is likely to 
be affected by scattering. However, there is a possibility that the point 
source is associated with a very compact emission at the head of the PWN, 
though the size estimates makes this extremely unlikely. The synchrotron 
self-absorption in this scenario gives an accurate model of the spectral 
nature. The GPS spectra in each pulsar was modeled using the thermal free-free
absorption. The speculative estimation of the absorber properties using the 
thermal absorption model were largely consistent with known physical conditions
in PWN.

\section*{Acknowledgments}
We thank the anonymous referee for the comments which helped to improve the 
paper. We thank the staff of the GMRT who have made these observations 
possible. The GMRT is run by the National Centre for Radio Astrophysics of the 
Tata Institute of Fundamental Research. This research was partially supported 
by the grant DEC-2013/09/B/ST9/02177 of the Polish National Science Centre.

\end{document}